\documentclass[referee]{raa}
\usepackage{graphicx,times}
\usepackage{natbib}
\usepackage{amssymb,amsmath}
\bibpunct{(}{)}{;}{a}{}{,}

\usepackage[a4paper=true,dvipdfm=true,pagebackref=true]{hyperref}
\hypersetup{pdftitle = The title of my PDF, pdfauthor = My name, pdfsubject= The subject, pdfkeywords = keyword1 keyword2 keyword3}
\hypersetup{colorlinks = true, linkcolor = green, anchorcolor = red, citecolor = blue, filecolor = red, pagecolor = red, urlcolor = red}

\begin{document}

\def\be{\begin{equation}}
\def\ee{\end{equation}}
\def\ba{\begin{eqnarray}}
\def\ea{\end{eqnarray}}
\def\trh{T_{\rm{RH}}}

   \title{Imprints of relic gravitational waves on pulsar timing$^*$
\footnotetext{\small $*$ Supported by National
Natural Science Foundation of China}
}

 \volnopage{ {\bf 2012} Vol.\ {\bf X} No. {\bf XX}, 000--000}
   \setcounter{page}{1}

   \author{Ming-Lei Tong\inst{1,2,3}, Yong-Heng Ding\inst{1,4}, Cheng-Shi Zhao\inst{1,2}, Feng Gao\inst{1,4}, Bao-Rong Yan\inst{1,5}, Ting-Gao Yang\inst{1,2},  Yu-Ping Gao\inst{1,2}
   }

   \institute{ National Time Service Center, Chinese Academy of Sciences, Xi'an 710600;
    {\it mltong@ntsc.ac.cn}\\
        \and
             Key Laboratory of Time and Frequency Primary Standards,
Chinese Academy of Sciences, Xi'an 710600\\
	\and 
	Key Laboratory for Researches in Galaxies and Cosmology, Chinese Academy of Sciences,  Hefei 230026\\
	\and 
	University of Chinese Academy of Sciences, Beijing 100049\\
	\and
	  Key Laboratory of Precision Navigation and Timing Technology,
Chinese Academy of Sciences, Xi'an 710600\\
\vs \no
   {\small Received xxx; accepted xxx}
}

\abstract{ Relic gravitational waves (RGWs) , a background originated  during inflation, would give imprints on the pulsar timing residuals. This makes RGWs be one of important sources for detection using the method of pulsar timing. In this paper, we discuss the effects of RGWs  on the single pulsar timing,  and give quantitively the timing residuals caused by RGWs with different model parameters. In principle, if the RGWs are strong enough today, they can be detected by timing  a single millisecond pulsar  with high precision  after the intrinsic red noise in pulsar timing residuals were understood, even though observing simultaneously  multiple millisecond pulsars is a more powerful technique in extracting gravitational wave signals.   We corrected the normalization of RGWs using observations of the cosmic microwave background (CMB), which leads to the  amplitudes of RGWs being  reduced by   two orders of magnitude or so compared to our previous works.  We made new constraints on RGWs using  the recent observations from the Parkes Pulsar Timing Array, employing  the tensor-to-scalar ratio $r=0.2$ due to the tensor-type polarization observations of CMB by BICEP2 as a referenced value even though it has been denied. Moreover, the constraints on RGWs from CMB and BBN (Big Bang nucleosynthesis)  will also be discussed for comparison. 
}

   \authorrunning{M.-L. Tong et al. }            
   \titlerunning{Imprints of relic gravitational waves on pulsar timing in light of BICEP2}  
   \maketitle

%
\section{Introduction}           
\label{sect:intro}

A stochastic background of relic
gravitational waves (RGWs) is predicted by the validities of  both general relativity and quantum mechanics (\citealt{Grishchuk+1975}; \citealt{Grishchuk+2001}; \citealt{Starobinsky+1979}; \citealt{Maggiore+2000}; \citealt{Zhang+etal+2005}; \citealt{Zhang+etal+2006}; \citealt{Miao+Zhang+2007}; \citealt{Giovannini+2010}). 
It was originated from the quantum fluctuations during the inflationary stage.
Hence, RGWs carry an unique information of the very early universe
and serve as a probe into the universe much earlier than  the cosmic microwave background (CMB) can do. 
As an advantage for detection, RGWs  spread a very broad frequency band, $\sim10^{-19}- 10^{10}$ Hz,
which make them be one of the major scientific targets  of various types of gravitational wave (GW) detectors,
such as the ground-based interferometers at $10^2-10^3$ Hz (\citealt{The LIGO Scientific Collaboration+The  Virgo Collaboration+2009}; \citealt{Willke+etal+2002}; \citealt{Acernese+etal+2005}; \citealt{Somiya+2012}),
the space-based interferometers at $10^{-4}-10^{-1}$ Hz ( \citealt{Seto+etal+2001}; \citealt{Crowder+Cornish+2005}; \citealt{Cutler+Harms+2006}; \citealt{Kawamura+etal+2006}; \citealt{Amaro-Seoane+etal+2012}), and high-frequency GW detectors around 100 MHz (\citealt{Cruise+2000}; \citealt{Tong+Zhang+2008}; \citealt{Li+etal+2003}; \citealt{Li+etal+2008}; \citealt{Tong+etal+2008}; \citealt{Akutsu+etal+2008}). 
For the  very low frequency around $10^{-18}$ Hz, RGWs can be detected by the measuring  the magnetic type of polarization of CMB (\citealt{Zaldarriaga+Seljak+1997}; \citealt{Kamionkowski+etal+1997}), which has been  a detecting goal of  WMAP (\citealt{Page+etal+2007}; \citealt{Komatsu+etal+2011}; \citealt{Hinshaw+etal+2013}),
Planck (\citealt{Planck Collaboration+2014}),  and BICEP2 (\citealt{Ade+etal+2014}).

Another important tool to detect RGWs directly is pulsar timing. The existence of a stochastic gravitational wave (GW) background will make  the  times of arrival (TOA) of the   pulses emitted from pulsars fluctuate.  The fluctuations of TOA are implied in the pulsar timing residuals.  
If multiple millisecond pulsars are observed simultaneously, forming the pulsar timing arrays (PTAs) (\citealt{Detweiler+1979}; \citealt{Romani+Taylor+1983}; \citealt{Hellings+Downs+1983}; \citealt{Kaspi+etal+1994}),   the GW signals can be extracted by correlating the timing residuals of each pair (\citealt{Detweiler+1979}; \citealt{Jenet+etal+2005}). PTAs respond to the frequency range of $10^{-9}-10^{-6}$ Hz, determined by the observational  Characteristics. Currently, there are several such projects running, 
such as the Parkes Pulsar Timing Array (PPTA) (\citealt{Hobbs+2008}; \citealt{Manchester+2013}),   
European Pulsar Timing Array (EPTA) (\cite{van Haasteren+etal+2011}),
the North American Nanohertz Observatory for Gravitational Waves (NANOGrav) (\citealt{Demorest+etal+2013}). The  much more sensitive Five-hundred-meter Aperture Spherical Radio Telescope (FAST) (\citealt{Nan+etal+2011}; \citealt{Hobbs+etal+2014}) and  Square Kilometre Array (SKA) (\citealt{Kramer+etal+2004}; \citealt{Janssen+etal+2015})
are also under planning.

Even though there is no direct detection of RGWs so far, one can still give constraints on RGWs based on the current observations and some conceivable theories. These  constraints could  prevent us from choosing some unreasonable parameters of RGWs. At present, various constraints on GW background have been studied.
The  recent observations from Parkes Pulsar Timing Array (PPTA) gave an upper limit on the energy density spectrum  $\Omega_g(f_{\rm PPTA})$ at $f_{\rm PPTA}=2.8$ nHz (\citealt{Shannon+etal+2013}).  On the other hand,  the physical  processes happened in the early universe can also given constraints on RGWs.   The successful Big Bang nucleosynthesis (BBN)
puts a tight upper bound on
the total energy fraction (\citealt{Maggiore+2000}) of GW background for frequencies $f>10^{-10}$ Hz (\citealt{Allen+Romano+1999}). 
The CMB  and matter power spectra 
also  give an upper limit of the total energy fraction of GWs
for   frequencies $f>10^{-15}$ Hz (\citealt{Smith+etal+2006}). Therefore, a successful model of RGWs should be compatible with the above constraints. 
 The RGWs  spectrum is,  to a large extent,
mainly described  by the initial amplitude normalized by the tensor-to-scalar ratio $r$ (\citealt{Boyle+Steinhardt+2008})  and 
the inflation spectral index  $\beta$ (\citealt{Grishchuk+2001}; \citealt{Zhang+etal+2005}; \citealt{Miao+Zhang+2007}; \citealt{Tong+Zhang+2009}). Here, we assume a zero  running spectral index $\alpha_t$ (\citealt{Tong+Zhang+2009}), since it only affects the spectrum at high frequencies. 
Since $\beta$ describes directly  the expansion behavior of inflation, the determination or constraint of $\beta$
would be  powerful in discriminating
inflationary models.
The afore-mentioned bounds  can be  converted into
the constraints on $\beta$  for a fixed $r$
(\citealt{Tong+Zhang+2009}; \citealt{Zhang+etal+2010})  and for a varying $r$ as a general analysis (\citealt{Tong+etal+2014}).
The WMAP observations of the spectra of CMB anisotropies and polarization
have yielded upper bounds on
the  ratio  $r$ of  RGWs
for the fixed scalar index (\citealt{Komatsu+etal+2011}).
Moreover, the recent observations of the polarizations of CMB by BICEP2 (\citealt{Ade+etal+2014})   gave a best estimation: $r=0.2$. Even though this result is  disfavored now,  we still use $r=0.2$ as a referenced value throughout this paper.

 It is worth to point out that, we corrected the normalization of the amplitude of RGWs at a pivot $k_0$  by using the tensor-type power spectrum of CMB at the time of $k_0$ mode re-entering the Hubble horizon instead of that at the 
 present time. This will reduce the amplitudes of RGWs for all the modes by two orders of magnitude or so compared to our previous works (\citealt{Tong+2012}; \citealt{Tong+etal+2014}). Hence, we will give new constraints on RGWs based on the theoretical RGWs and the updated observational data of PPTA. 
 As a general discussion, we will not employ the quantum normalization (\citealt{Grishchuk+2001}; \citealt{Tong+etal+2014})  in this paper.
 Based on that, we will study
 how RGWs affect pulsar timing with different values of $\beta$  both in the time domain and the frequency domain.  Moreover, we will quantitively calculated the corresponding pulsar timing residuals induced by RGWs for different $\beta$.  Even though observing simultaneously  multiple millisecond pulsars is a more powerful technique in extracting gravitational wave signals, in this paper we only discuss the effect of RGWs on the TOAs of an individual pulsar  since it is the basement for gravitational wave detection by PTAs.   In principle,  one can also extract  the signal of GWs from a single timing residuals with
 the assumption that the intrinsic red noises are understood. Thus, a comparison between    the  detecting sensitive curve determined by the 
ground clock and white-timing noise with the theoretical  spectra of RGWs will   be given.

Throughout  this paper we the use units in which the light speed  $c=1$.

\section{Relic gravitational waves in the accelerating universe}

In a spatially flat universe, the perturbed  Friedmann-Robertson-Walker metric under
the existence of the RGWs is
\begin{equation}
ds^2=a^2(\tau)[-d\tau^2+(\delta_{ij}+h_{ij})dx^idx^j],
\end{equation}
where $a(\tau)$ is the scale factor, $\tau$ is the conformal time, and $h_{ij}$ stands for the  perturbations to the  homogenous and isotropic spacetime background due to RGWs. According to Einstein field equation, RGWs satisfy
\begin{equation}\label{weq}
\partial_\mu(\sqrt{-g}\partial^\mu h_{ij}(\tau,{\bf{x}}))=0,
\end{equation}
where $g\equiv {\rm det}(g_{\mu\nu})$.
The general solution of Eq. (\ref{weq}) can be expanded as the Fourier  $k$-modes space, and has the following form:
\begin{equation}
\label{planwave}
h_{ij}(\tau,{\bf x})=
   \sum_{A=+,\times}\int\frac{d^3\bf{k}}{(2\pi)^{3/2}}
         \epsilon^{A}_{ij}h_k^{A}(\tau)e^{i\bf{k}\cdot{x}},
\end{equation}
where $A$ stands for the two polarization states, under the transverse-traceless (TT) gauge. Since the two polarizations of
$h^{A}_k(\tau )$ have the same statistical
properties and give equal contributions to the unpolarized RGWs background,
  the super index $A$ can be dropped.  For a power-law form of $a(\tau) \propto \tau^\alpha$,
 $h_k(\tau )$ has an analytic
solution which is a linear combination of
Bessel and Neumann functions (\citealt{Zhang+etal+2005}; \citealt{Zhang+etal+2006}; \citealt{Miao+Zhang+2007}). 
In fact, the scale factor in all the cosmic expansion stages of the universe  can be written
in power-law forms (\citealt{Grishchuk+2001}; \citealt{Miao+Zhang+2007}; \citealt{Tong+Zhang+2009}; \citealt{Tong+2012}). For example, 
the scale factor in the inflationary stage has the following form: 
\begin{equation} \label{inflation}
a(\tau)=l_0|\tau|^{1+\beta},\,\,\,\,-\infty<\tau\leq \tau_1,
\end{equation}
where the inflation index $\beta$
is a model parameter describing the expansion behavior of the inflation, and $\tau_1$ denotes the end of the inflation. 
The special case of $\beta=-2$  corresponds  to the exact de Sitter expansion
driven by a constant vacuum energy density.
However, for inflationary expansions driven by some dynamic field,
the predicted values of $\beta$ could  deviate from   $-2$,
depending on specific models.
In the single-field
slow-roll inflation model, one always has $\beta<-2$, i.e., red spectrum (\citealt{Liddle+Lyth+2000}).
 For example, the relation $n_s=2\beta+5$ which is often employed (\citealt{Grishchuk+2001}; \citealt{Tong+Zhang+2009})
 gives $\beta=-2.02$ for $n_s\simeq0.96$ based on the observation of CMB by Planck (\citealt{Planck Collaboration+2014}). However, some other inflation
models, such as the phantom inflations (\citealt{Piao+Zhang+2004}) also
predict the blue spectrum, which has not been excluded
by observations (\citealt{Stewart+Brandenberger+2008}; \citealt{Camerini+etal+2008}).
Below, we recognize $\beta$ as a major free parameter of RGWs.
As shown in \citealt{Tong+2013}, 
$\beta_s$ describing the expansion behavior of the reheating process only affects the RGWs in very high frequencies which are far above the upper limit
frequency of the pulsar timing response.
In this paper, we will take  $\beta_s=1$ (\citealt{Starobinsky+1980}; \citealt{Kuroyanagi+etal+2009}).
After the  radiation-dominant stage and 
the matter-dominant stage, the universe is undergoing an accelerating stage, where the scale factor has 
the following form: 
\begin{equation} \label{accel}
a(\tau)=l_H|\tau-\tau_a|^{-\gamma},
 \end{equation}
where $\gamma\simeq 3.5$ can be determined by numerically fitting method with  the energy density contrast
$\Omega_{\Lambda}=0.685$ given by Planck+WMAP (\citealt{Planck Collaboration+2014}).  
Conveniently,
  $|\tau_0-\tau_a|=1$  was employed (\citealt{Zhang+etal+2005}; \citealt{Zhang+etal+2006}), i.e.,
the present scale factor $a(\tau_0)=l_H$.
By  definition,
one has $l_H=\gamma/H_0$,
where the Hubble constant $H_0=100\, h$ km s$^{-1}$ Mpc$^{-1}$
with  $h= 0.673$  (\citealt{Planck Collaboration+2014}).
The coefficients and constants embedded in the expressions of the scale factors
 can be determined by the continuity of $a(\tau)$ and $a'(\tau)$
at the   points  joining the various stages. 

The  increases of
the scale factor for different stages are defined as:
$\zeta_1\equiv{a(\tau_s)}/{a(\tau_1)}$,
$\zeta_s\equiv{a(\tau_2)}/{a(\tau_s)}$,
$\zeta_2\equiv{a(\tau_E)}/{a(\tau_2)}$,
and
$\zeta_E\equiv{a(\tau_0)}/{a(\tau_E)}$, where $\tau_s$, $\tau_2$ and $\tau_E$ represent the beginning of the radiation-dominated stage, the matter-dominated stage and the accelerating stage, respectively. 
For the accelerating stage in the simple $\Lambda$CDM model,
one  has
$\zeta_E =1+z_E \simeq ({\Omega_\Lambda}/{\Omega_m})^{1/3}$,
where $z_E$ is the redshift when the  accelerating expansion begins.
For the matter-dominated stage,
one has  $\zeta_2 
=(1+z_{eq}) \zeta_E^{-1}$ with $z_{eq}=3402$  (\citealt{Planck Collaboration+2014}).
For the radiation-dominated stage, the value of $\zeta_s$ depends on
the reheating temperature $\trh$,
at which the radiation-dominated stage begins. Due to the conservation of the entropy, $\zeta_s$ can be written in terms of $\trh$ (\citealt{Tong+2012}; \citealt{Tong+2013}):
\begin{equation}\label{delta1}
\zeta_s=\frac{T_{\rm{RH}}}{T_{\rm{CMB}}(1+z_{eq})}
          \left(\frac{g_{\ast s}}{g_{\star s}}\right)^{1/3},
\end{equation}
where   $T_{\rm{CMB}}=2.725\, {\rm K} =2.348\times10^{-13}$ GeV is the
present CMB temperature,
$g_{\ast s}\simeq200$ is the effective number of relativistic species contributing
to the entropy after the reheating,
and  $g_{\star s}=3.91$ is the one after recombination (\citealt{Watanabe+Komatsu+2006}; \citealt{Tong+2012}).
For the single field inflation, CMB data would yield
 the lower bound of $\trh \gtrsim 6\times10^3$ GeV,
and the most upper bound  could be up
to  $\trh \lesssim3\times10^{15}$ GeV  (\citealt{Martin+Ringeval+2010}).
Some model like the slow-roll massive scalar field inflation predicts a definite value  $\trh = 5.8\times 10^{14}$ GeV (\citealt{Tong+2012}).
In this paper we generally
 consider  a large  range of  $\trh \sim(10^4-10^{15})$ GeV for a complete demonstration.
The uncertainty of $\trh$ is due to the lack of knowledges of 
the reheating process happened following the inflationary expansion that
converts the vacuum energy into radiation. 
So 
the parameter  $\zeta_1$ is also uncertain.
Based on the slow-roll scalar inflation models (\citealt{Mielczarek+2011}; \citealt{Tong+2012}; \citealt{Tong+2013}),
 $\zeta_1$ depends on the specific form of the potential $V$ that drives the inflation. However, the determination of $\zeta_1$ in that method has a very large relative uncertainty. 
If we  calculate the spectra of RGWs in low frequencies,  some particular values of $\zeta_1$ can be set as it only affects RGWs in very high frequencies.

The spectrum of RGWs $h(k,\tau)$ is defined by
\begin{equation} \langle
h^{ij}(\tau,\mathbf{x})h_{ij}(\tau,\mathbf{x})\rangle\equiv\int_0^\infty
h^2(k,\tau)\frac{dk}{k},
\end{equation}
where the angle brackets mean ensemble average. The present RGWs spectrum  relates to  the {\it characteristic strain spectrum} (\citealt{Maggiore+2000}) or {\it chirp amplitude} (\citealt{Boyle+etal+2006}; \citealt{Boyle+Steinhardt+2008}) as $h_c(f)\equiv h(f,\tau_0)/\sqrt{2}$.
Assuming that the wave mode crosses the  horizon of the universe when
$\lambda/(2\pi)=1/H$, then the characteristic comoving wave number
at a certain joining time $\tau_x$ can be defined as (\citealt{Tong+etal+2014})
\begin{equation}\label{wavenumber}
k_x\equiv k(\tau_x) =  a(\tau_x) H(\tau_x).
\end{equation}
 For example, the characteristic comoving wave number at present is
 $k_H=a(\tau_0)H_0=\gamma$. By a similar calculation,
one has  the following relations:
\begin{equation}\label{frelation}
  \frac{k_E}{k_H}
    = \zeta_E^{-\frac{1}{\gamma}},\quad
    \frac{k_2}{k_E}=\zeta_2^{\frac{1}{2}},\quad
    \frac{k_s}{k_2}=\zeta_s, \quad
    \frac{k_1}{k_s}=\zeta_1^{\frac{1}{1+\beta_s}}.
 \end{equation}
In the present universe,
the physical frequency relates to a comoving wave number  $k$ as
\begin{equation} \label{freq}
f=  \frac{k}{2\pi a (\tau_0)} = \frac{k}{2\pi l_H}.
\end{equation}
Thus, one can easily has $f_H=H_0/2\pi$, and other characteristic frequencies can be easily determined subsequently by Eq. (\ref{frelation}). Note that,  $f_s$ depends on the value of $\trh$.
The present energy density contrast of RGWs  is defined by
$ \Omega_{GW}=\langle\rho_{g}\rangle/{\rho_c}$,
where $\rho_g$
is the energy density of RGWs
and $\rho_c=3H_0^2/8\pi G$ is the critical energy density.
The dimensionless  energy density spectrum  relates to the characteristic amplitude of 
RGWs as (\citealt{Grishchuk+2001}; \citealt{Maggiore+2000})
 \begin{equation}\label{omega}
\Omega_g(f)=\frac{d\Omega_{GW}}{d\ln{f}}=\frac{2\pi^2}{3}
        h^2_c(f)
     \Big(\frac{f}{H_0}\Big)^2.
\end{equation}
The analytic solutions of the RGWs were studied by many authors
(\citealt{Zhang+etal+2006}; \citealt{Watanabe+Komatsu+2006}; \citealt{Miao+Zhang+2007}; \citealt{Kuroyanagi+etal+2009}; \citealt{Tong+Zhang+2009}). On the other hand,
 the approximate solutions of RGWs in the whole frequency range were listed in \citealt{Tong+2012}. 
Even though the initial amplitude of RGWs can be, in principle,  given by the quantum normalization condition (\citealt{Tong+etal+2014}),  it relies on many physical processes which  are not well known so far. Here, for a simple discussion, we will not consider the normalization condition. On the other hand, the initial amplitude should be normalized to observations.  Below, we will determine the initial amplitude of RGWs due to the observations of CMB.

\section{Constraints on RGWs by current observations}

\subsection{Determine the primordial amplitude of RGWs from observations of CMB}

From the observations of the B-mode polarization in  the spectrum of CMB, the power spectrum  of RGWs
at a pivot wave number $k_0/a(\tau_0)=0.002$ Mpc$^{-1}$ can be normalized to the scalar power spectrum using the tensor-to-scalar ratio (\citealt{Peiris+etal+2003}; \citealt{Spergel+etal+2007}; \citealt{Komatsu+etal+2011}):
\begin{equation}\label{ratio}
r\equiv\frac{\Delta^2_h(k_0)}{\Delta^2_{\mathcal{R}}(k_0)},
\end{equation}
where $\Delta^2_h(k_0)\equiv h^2(k_0,\tau_i)$ (\citealt{Boyle+Steinhardt+2008}) with  $\tau_i$ denoting the  moment of  a mode $k$  re-enters the Hubble horizon, and the scalar power spectrum $\Delta^2_{\mathcal{R}}(k_0)=2.427\times10^{-9}$ given
by WMAP\,9+BAO+$H_0$ (\citealt{Hinshaw+etal+2013}). It is worth to point out that we made a wrong normalization in our previous work (\citealt{Tong+2012}; \citealt{Tong+etal+2014}), where   $\Delta^2_h(k_0)\equiv h^2(k_0,\tau_0)$ was employed. It will  overestimate the  spectrum of RGWs by two orders of magnitude  or so, which will be analyzed later. 
 Recently,  
  the detection of B-mode polarization at degree angular scales in CMB by BICEP2 (\citealt{Ade+etal+2014})  gave a definite
  value $r=0.2^{+0.07}_{-0.05}$. Even though this result is denied (\citealt{BICEP2/Keck and Planck Collaborations+2015}),
we still  take $r=0.2$ in the following for an  tentative demonstration.  Besides, CMB observations can also give constraints on the ratio $r$ (\citealt{Hinshaw+etal+2013}; \citealt{Planck Collaboration+2014}; \citealt{BICEP2/Keck and Planck Collaborations+2015}).  For example, \citealt{Planck Collaboration+2014} gave $r<0.11$ and $r<0.26$ for
  the  vanishing $\alpha_s$ and the  non-vanishing $\alpha_s$, respectively.    The current limit is given by a joint analysis of BICEP2/Keck Array and Planck Data (\citealt{BICEP2/Keck and Planck Collaborations+2015}):  $r_{0.05}<0.12$ at $95\%$ confidence level.   On the other hand,  a lower limit $r\gtrsim10^{-2}$ was obtained  (\citealt{Boyle+etal+2006}) using a discrete,
model-independent measure of the degree
of fine-tuning required, if $0.95\lesssim n_s<0.98$,
in accord with current measurements.

According to the approximate solutions listed in  \citealt{Tong+2012}, the spectrum of RGWs at $f=f_0$ satisfies
\be\label{hnorm}
h(k_0,\tau_0)=A\left(\frac{f_0}{f_H}\right)^{\beta}(1+z_E)^{-\frac{2+\gamma}{\gamma}},
\ee
where $A$ stands for  the initial amplitude of RGWs and it can be determined by observations. Eq. (\ref{hnorm}) means  that  $k_0$-mode of RGWs re-entered the horizon at the matter-dominated stage since $f_H<f_0<f_2$. So $h(k_0,\tau_0)$ has been suffered a decay from the $k_0$-mode re-entered the horizon to the present time, i.e., $h(k_0,\tau_0)=h(k_0,\tau_i) \frac{a(\tau_i)}{a(\tau_0)}$, where 
$\frac{a(\tau_i)}{a(\tau_0)}$ is the decaying factor.  From Eqs. (\ref{frelation}) and (\ref{hnorm}), one can easily have
\begin{equation}\label{hki}
h(k_0,\tau_i)=A\left(\frac{f_0}{f_H}\right)^{2+\beta}.
\end{equation}
Combining Eqs. (\ref{ratio}) and (\ref{hki}),  one has
\begin{equation}\label{primordial}
A 
=\sqrt{\Delta^2_{\mathcal{R}}(k_0)r}\left(\frac{f_H}{f_0}\right)^{2+\beta}.
\end{equation}
Hence,  $A$ can be determined
for the given   $r$ and  $\beta$.  Figure 1 shows a comparison of the analytic spectrum and the 
approximate spectrum of RGWs for   $\trh=10^{15}$ GeV,  where $\beta=-2$ and $\zeta_1=10^5$ were also set. The value of $\zeta_1$ was chosen that  $f_1$ should be lower than the upper limit frequency $\sim4\times10^{10}$ Hz (\citealt{Grishchuk+2001}; \citealt{Tong+2012}). One can see that the approximate spectrum is in accord with the analytic spectrum very well. So we  can constrain some model parameters using the approximate spectra of RGWs simply from  observations.  
 $\zeta_s$ depends on $\trh$ linearly, so dose $f_s$ as can be seen from Eq. (\ref{frelation}).
We plotted the approximate spectrum  for the cases of $\trh=10^4$ GeV in Figure 1.  
It is  clear that different $\trh$ only affects the spectrum in the very high frequencies ($\gtrsim10^{-3}$ Hz).  Since pulsar timing arrays respond to the frequencies localized at the range of $10^{-9}-10^{-6}$ Hz, the imprints of RGWs on  the  pulsar timing arrays   do not depend on the value of $\trh$.

\begin{figure}
   \centering
   \includegraphics[width=12.0cm, angle=0]{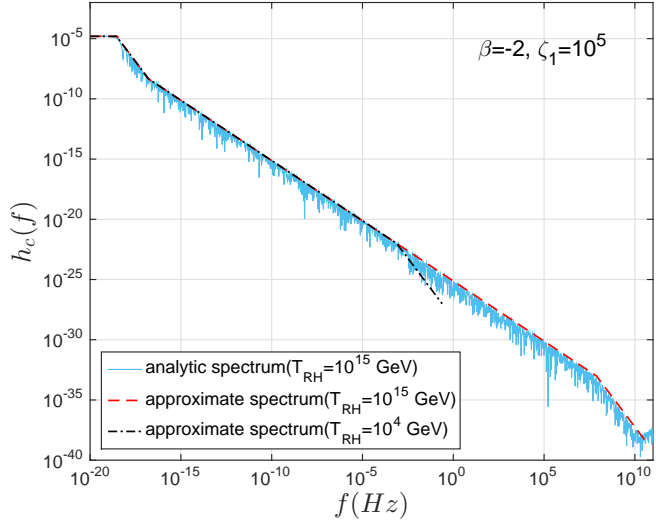}
   \caption{The comparison between the analytic (solid line) and approximate (dashed line) spectra of RGWs with a fixed parameter set of $\beta=-2$, $\zeta_1=10^{-5}$ and $\trh=10^{15}$ GeV. The approximate spectrum (dot-dashed line) of RGWs with   $\trh=10^4$ GeV was also plotted for comparison. }
   \label{Fig1}
   \end{figure}

\subsection{Constraints on $\beta$ by pulsar timing arrays and the very early universe}

 PTA experiments have set constraints on GW background (\citealt{Bertotti+etal+1983}; \citealt{Kaspi+etal+1994};\citealt{Thorsett+Dewey+1996}; \citealt{McHugh+etal+1996}; \citealt{Jenet+etal+2006}; \citealt{Hobbs+etal+2009}; \citealt{van Haasteren+etal+2011}; \citealt{Demorest+etal+2013}; \citealt{Zhao+etal+2013}).
In the  data analysis of PTAs,
the characteristic strain spectrum of GW background is usually modeled   with a power-law form:
 \be\label{hc1}
 h_c(f)=h_{{\rm yr}}\left(\frac{f}{{\rm yr}^{-1}}\right)^\alpha,
 \ee
where $h_{yr}$ is the amplitude at $f={\rm yr}^{-1}$.
For the frequency band  $10^{-9}\leq f\leq10^{-6}$ Hz of PTA   experiments, the 
 characteristic strain spectrum 
has the following form (\citealt{Tong+2012}):
\be\label{chara}
h_c(f)=\frac{A}{\sqrt{2}}
\left(\frac{f}{f_H}\right)^{1+\beta}
\left(\frac{f_H}{f_2}\right)(1+z_E)^{-\frac{2+\gamma}{\gamma}}.
\ee
With the help of Eq. (\ref{primordial}),
Eq. (\ref{chara}) can be rewritten as
\be\label{strain}
h_c(f)=\sqrt{\frac{\Delta_\mathcal{R}^2(k_0)r}{2}}
\left(\frac{f_H^2}{f_0f_2}\right)
\left(\frac{f}{f_0}\right)^{1+\beta}(1+z_E)^{-\frac{2+\gamma}{\gamma}}.
\ee
Note that $f/f_0\gg1$ in the pulsar timing frequency band.
Comparing  Eq.(\ref{hc1}) and Eq.(\ref{strain}) tells that
the power-law index is related to the inflation index via
\be
\alpha=1+\beta.
\ee

Improving the earlier  work (eg. \citealt{Kaspi+etal+1994}), \citealt{Jenet+etal+2006}
  developed a frequentist technique of statistics,
and have placed an upper limit on $h_{yr}$ for different values of  $\alpha$.
Recently, \citealt{Shannon+etal+2013}  gave an upper limit $h_{yr}<2.4\times10^{-15}$ at the $95\%$ confidence level for $\alpha=-2/3$ using data from PPTA and that available observations from the Arecibo observatory.  Even though this limit is intended for supermassive black hole binaries, one can equivalently translate it to the case of  RGWs, which leads to $h_{yr}<1.0\times10^{-15}$ for $\alpha=-1$.  Note that these limits are independent of $H_0$.  
Fig. 2 gives the upper limit curves of $h_{yr}(\alpha)$  for PPTA at different phases. The PPTA(2006) and the future PPTA obtained from the simulated data of the potential 20 pulsars for the future goal of the PPTA timing are taken from \citealt{Jenet+etal+2006}. To constrain the parameter $\alpha$, $h_c(\rm{yr}^{-1})$ with $r=0.2$ was also plotted. 
As shown in \citealt{Tong+etal+2014}, the condition $h_c({\rm yr}^{-1})<h_{yr}(\alpha)$ leads to an upper limit of $\alpha$ (or $\beta$) for a given $r$. Since $r=0.2$ is set in this paper,  the PPTA(2013) gives an
upper limit $\alpha<-0.70$ ($\beta<-1.70$). Comparably, PPTA(2006) gives an upper limit $\alpha<-0.63$
 ($\beta<-1.63$), and the future PPTA will give a limit $\alpha<-0.76$ ($\beta<-1.76$).

\begin{figure}
   \centering
   \includegraphics[width=12.0cm, angle=0]{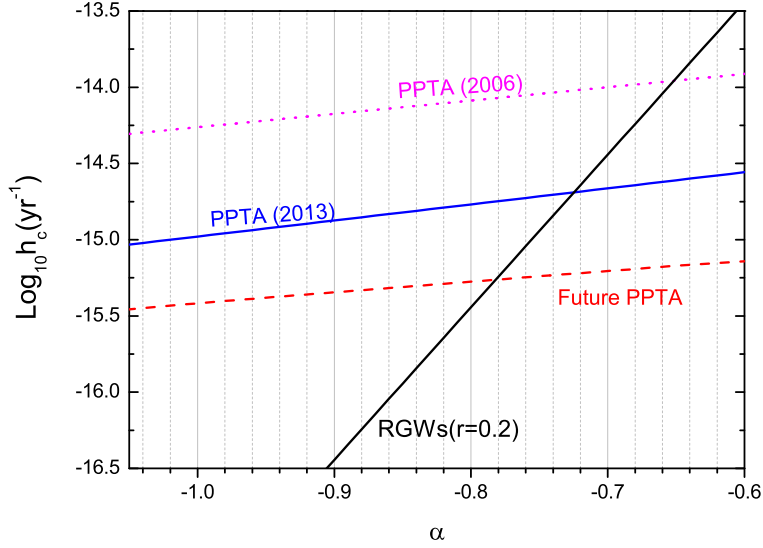}
   \caption{The constraints of $\alpha$ given by the PPTA(2006) (\citealt{Jenet+etal+2006}), the PPTA(2013) (\citealt{Shannon+etal+2013}) and the future PPTA (\citealt{Jenet+etal+2006}), respectively, confronting RGWs with $r=0.2$. These constraints are made at the frequency of one cycle per year. }
   \label{Fig2}
   \end{figure}

   Besides the constraints from PTAs,  some other observations also give constraints on RGWs. For example, $\beta$ can be  constrained by ground-based laser interferometer (\citealt{Tong+Zhang+2009}; \citealt{Chen+etal+2014}). However, the ground-based laser interferometers respond to RGWs at the frequency range of $10^2-10^3$ Hz, and the RGWs with high frequencies  depend on theoretical models which are not known well. For instance, if we choose $\trh=10^4$ GeV, there will be no RGWs  with frequencies lager than $0.24$ Hz. In addition, BBN and CMB  can give constraints on the GW energy density contrast at the time of nucleosynthesis and CMB decoupling, respectively.  The constraint from BBN is given by $\Omega_{GW}^{\rm{BBN}}<1.1\times10^{-5}(N_\nu-3)$ (\citealt{Maggiore+2000}), where the effective number of neutrino species at the time of BBN has an upper bound $N_\nu-3<1.4$ (\citealt{Cyburt+etal+2005}). On the other hand, CMB gives  $\Omega_{GW}^{\rm{CMB}}<1.3\times10^{-5}$ (\citealt{Smith+etal+2006}). Note that, the lower frequency limits   contributing the energy density contrasts for BBN and CMB are different. For BBN $f_{low}\sim10^{-10}$ Hz,  corresponds to the horizon scale at the time of BBN (\citealt{Allen+Romano+1999}); while for CMB $f_{low}\sim10^{-15}$,   corresponds  to the horizon scale at the decoupling of CMB (\citealt{Zhang+etal+2010}). However, the upper limit frequency for the two cases is both $f_1$.  As pointed out above, $f_1$ depends of $\zeta_s$, and thus depends of $\trh$ in turn. So the constraints on $\beta$ given by BBN or CMB depend on $\trh$. As shown in Eq. (\ref{frelation}), $f_1$ also depends on $\zeta_1$, however, we found that the limits of $\beta$ constrained by BBN/CMB are nearly independent on $\zeta_1$. It can explained that   
   $\zeta_1$ only affects the power spectrum in the frequency band ($f_s$, $f_1$), which contributes very little to the total energy density contrast.  It  was also demonstrated clearly in Fig. 1 of \citealt{Tong+etal+2014}.   Fig. 3 shows the upper limits of $\beta$ constrained by BBN and CMB with varying $\trh$. One can see that BBN and CMB almost give the same constraints for the whole range of $\trh$.  The most stringent constraint is given by $\beta<-1.83$ with $\trh=10^{15}$ GeV, which is more stringent than those given by the present PPTA data. 
 However, one should put it in mind that the constraints of $\beta$ given by PTAs are independent of $\trh$.

\begin{figure}
   \centering
   \includegraphics[width=10.0cm, angle=0]{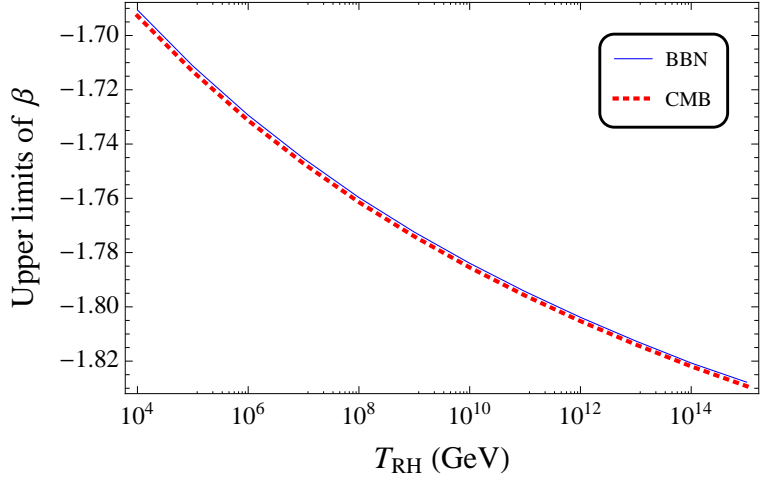}
   \caption{Upper limits of $\beta$ with varying $\trh$ given by BBN and CMB , respectively.}
   \label{Fig3}
   \end{figure}

\section{ Imprints of RGWs on pulsar timing}

The existence of gravitational waves
 will change the geodesic of the photons  from millisecond
 pulsars to the observer.
Consequently, the TOAs of the electromagnetic signals from pulsars
 will be perturbed, forming the so-called timing residuals if the
 effect of RGWs is not taken into account in the timing model.
 The  GW background would lead to a red power-law spectrum of the timing
 residuals. Even though such red spectra can also be produced by intrinsic noise of pulsars (\citealt{Verbiest+etal+2009}; \citealt{Shannon+Cordes+2010}), inaccuracies in the solar system ephemeris (\citealt{Champion+etal+2010}), or variations in terrestrial time standards (\citealt{Hobbs+etal+2012}),
a GW background produces a unique signature in the timing residuals
that can be confirmed by observing correlated signals between multiple pulsars
widely distributed on the sky (\citealt{Hellings+Downs+1983}; \citealt{Jenet+etal+2005}). 
However, here we only analyze how RGWs affect the timing residuals of a single pulsar. 
In principle, if the GW signals are strong enough,
one could extract their signals buried in
the data of the timing residual measurements after all known effects have been accounted for.
On the other hand, even GWs are very weak, one can still 
constrain  the amplitude of GWs 
with the long-time accumulating data of timing residuals.

The  frequency of the signals from a pulsar will be shifted due to the existence of a GW. For a GW 
propagating in the direction $\hat{\Omega}$, the redshift of the signals from a pulsar in the direction $\hat{p}$ is given by (\citealt{Detweiler+1979}; \citealt{Anholm+etal+2009})
\be\label{redshiftz}
z(t,\hat{\Omega})=\frac{\nu_e-\nu_p}{\nu_p}=\frac{\hat{p}^i\hat{p}^j}{2(1+\hat{\Omega}\cdot\hat{p})}\Delta h_{ij},
\ee
where $\nu_e$, $\nu_p$ represent the frequencies of the pulse received at the  Earth and the pulse emitted at the pulsar, respectively, and 
\be
\Delta h_{ij}=[h_{ij}(t_p,\hat{\Omega})-h_{ij}(t_e,\hat{\Omega})]
\ee
is the difference in the metric perturbation traveling along the direction $\hat{\Omega}$ at the pulsar and at the Earth. $t_p$ and $t_e$ are the times at which the GW passes the pulsar and the    Earth, respectively. Note that, the standard Einstein summation convention was used in Eq. (\ref{redshiftz}). 
 The vectors $(t_e,{\bf{x}}_e)$ and $(t_p,{\bf{x}}_p)$ give the spacetime coordinates of the Solar System barycenter (SSB) and the pulsar, respectively. In the following, we choose a coordinate system that the origin of the space coordinates is located at the SSB, and   use $t$ instead of $t_e$ denoting the time coordinate. Moreover, the following conventions are often used (\citealt{Anholm+etal+2009}),
\be
t_p=t-L, \quad {\bf{x}}_p=L\hat{p},
\ee  
where $L$ is the distance of the pulsar away from the SSB. If we assume that the amplitude of the GW is the same at the SSB and  the pulsar then $\Delta h_{ij}$ can be written as the following Fourier integration form (\citealt{Anholm+etal+2009}):
\be\label{deltah}
\Delta h_{ij}(t,\hat{\Omega})=\sum_{A=+,\times}\int_{-\infty}^{\infty}df e^{i2\pi f t}(e^{-i2\pi f L (1+\hat{\Omega}\cdot\hat{p})}-1) h_A(f,\hat{\Omega})\epsilon_{ij}^A(\hat{\Omega}),
\ee
where $A$ stands for the two polarizations of GWs, and  the corresponding tensor $\epsilon_{ij}^A(\Omega)$ can be written as 
\be
\epsilon_{ij}^+(\hat{\Omega})=\hat{m}_i\hat{m}_j-\hat{n}_i\hat{n}_j, \quad \epsilon_{ij}^{\times}(\hat{\Omega})=\hat{m}_i\hat{n}_j+\hat{n}_i\hat{m}_j,
\ee
with $\hat{m},\hat{n}$ unit vectors orthognal to $\hat{\Omega}$ and to each other. One has straightforwardly, 
\be
\epsilon_{ij}^A(\hat{\Omega})\epsilon^{A',ij}(\hat{\Omega})=2\delta^{AA'}.
\ee
If we assume that the  stochastic background is  isotropic, unpolarized and stationary, the ensemble average of the Fourier amplitudes can be written as
\be\label{hfourier}
\langle\tilde{h}_A^\ast(f,\hat{\Omega})\tilde{h}_{A'}(f',\hat{\Omega}')\rangle=\frac{1}{16\pi}\delta(f-f')\delta^2(\hat{\Omega},\hat{\Omega}')\delta_{AA'}S_h(f).
\ee
Note that, the spectral density $S_h(f)$ defined above is twice as much as  that shown in Eq.(8) of \citealt{Maggiore+2000}, and  satisfies $S_h(f)=S_h(-f)$. Here $S_h(f)$ is called {\it one-sided spectral density}, and it is related to the characteristic strain amplitude as (\citealt{Maggiore+2000})
\be
h_c^2(f)=fS_h(f).
\ee
Substitute Eq. (\ref{deltah}) into Eq.(\ref{redshiftz}), one has
\be\label{zomega}
 z(t,\hat{\Omega})=\sum_A\int_{-\infty}^{\infty}df e^{i2\pi f t}(e^{-i2\pi f L (1+\hat{\Omega}\cdot\hat{p})}-1) h_A(f,\hat{\Omega})F^A(\hat{\Omega}),
\ee
where 
\be\label{F}
F^A(\hat{\Omega})\equiv \epsilon_{ij}^A(\hat{\Omega})\frac{1}{2}\frac{\hat{p}^i\hat{p}^j}{1+\hat{\Omega}\cdot\hat{p}}
\ee
has been defined. For a stochastic GW background, the total redshift is given by summing over the contributions coming from GWs in every direction (\citealt{Anholm+etal+2009})
\be\label{zt}
z(t)=\int_{S^2}d\hat{\Omega}\  z(t,\hat{\Omega}).
\ee  
The pulsar timing residual is defined as the integral of the redshift 
\be
R(t)\equiv\int_0^t dt' z(t').
\ee

The total relative frequency changes can be divided into two parts:
\be
s(t)=z(t)+n(t),
\ee
where $z(t)$ is induced by the RGWs and $n(t)$ is the noise intrinsic in the timing measurement which is assumed  to be stationary and Gaussian. In addition, we also assume that 
\be
\langle z(t)\rangle=\langle n(t)\rangle=0, \quad \langle z(t) n(t)\rangle=0,
\ee
where the angle brackets denote an expectation value.  With the help of Eqs.(\ref{hfourier}), (\ref{zomega}) and (\ref{zt}), the variance of the relative frequency changes is given by 
\be
\langle z^2(t)\rangle\equiv\int_0^\infty df S_z(f)=F\int_0^\infty df S_h(f),
\ee
where $S_z(f)=FS_h(f)$ and 
\be
F\equiv \frac{1}{8\pi}\int_{S^2}d \hat{\Omega} \left|e^{i2\pi fL(1+\hat{\Omega}\cdot\hat{p})}-1\right|^2\sum_AF^A(\hat{\Omega})F^A(\hat{\Omega}).
\ee
Using the definition in Eq.(\ref{F}), one can easily obtain (\citealt{Jenet+etal+2011}) 
\be
F=\frac{1}{3}-\frac{1}{8\pi^2f^2L^2}+\frac{\sin{(4\pi fL)}}{32\pi^3f^3L^3}.
\ee
The property of $F$ is shown in Fig. 4. It can be seen that $F$ is generally frequency-dependent, however, F will converge to a fixed value $1/3$ when $fL\gtrsim3$. For  pulsar timing experiments, the distances of millisecond pulsars are usually larger than 0.1  kpc, and then $fL>10$ for the frequencies $f>10^{-9}$  Hz. Therefore, one always has $F=1/3$ for pulsar timing experiments. The factor $F$ represents the  root-mean-square (rms) signal response  averaged over the sky and polarization states.

\begin{figure}
   \centering
   \includegraphics[width=9.0cm, angle=0]{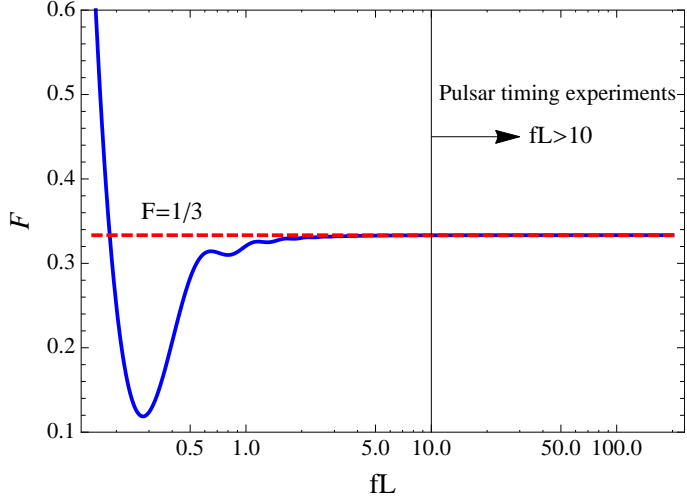}
   \caption{The property  of the reduction factor $F$ with different regimes of $fL$. For the pulsar timing experiments, one has $fL>10$. }
   \label{Fig4}
   \end{figure}

 Similarly, the ensemble average of the Fourier components of the noise satisfies
\be
\langle\tilde{n}^\ast(f)\tilde{n}(f')\rangle=\frac{1}{2}\delta(f-f')S_{n}(f),
\ee
where the noise spectrum density satisfies $S_{n}(f)=S_n(-f)$ with  dimension Hz$^{-1}$.
So, the variance of the noises is
\be
\langle n^2(t)\rangle=\int_0^\infty df S_{n}(f).
\ee 
Equivalently, the noise level of a GW detector is usually measured by the {\it strain sensitivity} $\tilde{h}(f)\equiv\sqrt{S_n(f)}$ with  dimension Hz$^{-1/2}$.  
For a given  signal-to-noise ratio (SNR),  one can discuss the ability of a detector to reach the minimum detectable amplitude of RGWs. 
Under the assumption that the dispersion caused by the interplanetary plasma is adequately calibrated and the intrinsic rotation instability of the pulsar can be negligible  or be well understood, the noise spectrum is characterized by the contribution due to the ground clock and a white-timing noise of 100 ns in a Fourier band $+/-0.5$ cycles/day (\citealt{da Silva Alves+Tinto+2011}). The 100 ns level is the current timing goal of PTAs and three pulsars are being timed to this level (\citealt{Verbiest+etal+2009}). Following the noise model discussed in \citealt{Jenet+etal+2011}, the expression for the noise spectrum density of the relative frequency fluctuations is (\citealt{da Silva Alves+Tinto+2011})
\be
S_{zn}(f)=[4.0\times10^{-31}f^{-1}+3.41\times10^{-8}f^2]\  {\rm{Hz}}^{-1}.
\ee
    For SNR=1, i.e., $S_z(f)/S_{zn}(f)=1$,  we plotted the strain sensitivity of the detection on GW background by a single pulsar and the analytic strain amplitude per root Hz (considering the $F$ factor) of RGWs, $h_c(f)\sqrt{F/f}$ (\citealt{Zhang+etal+2010}), with different values of $\beta$ in Fig. 5. One can see that it is hard to detect RGWs even for the blue spectrum by an individual pulsar timing method,  however, the lower frequencies of RGWs are more hopeful to be detected. Note that,  SNR=5 is conventionally taken as detection threshold for PTAs. Thus, the sensitivity curve shown in Fig. 5 should be multiplied a factor of $\sqrt{5}$. Therefore,  significant SNR improvements of 
pulsar timing sensitivities of radio
 telescopes will be required for reliable detection. 
  There are two ways for this  target. First, one can simultaneously timing several pulsars. The SNR can be improved by correlating the data of several pulsars just as the method used in the networks of ground-based interferometers. Second, one should try to suppress the various  noises in the timing measurement.

\begin{figure}
   \centering
   \includegraphics[width=12.0cm, angle=0]{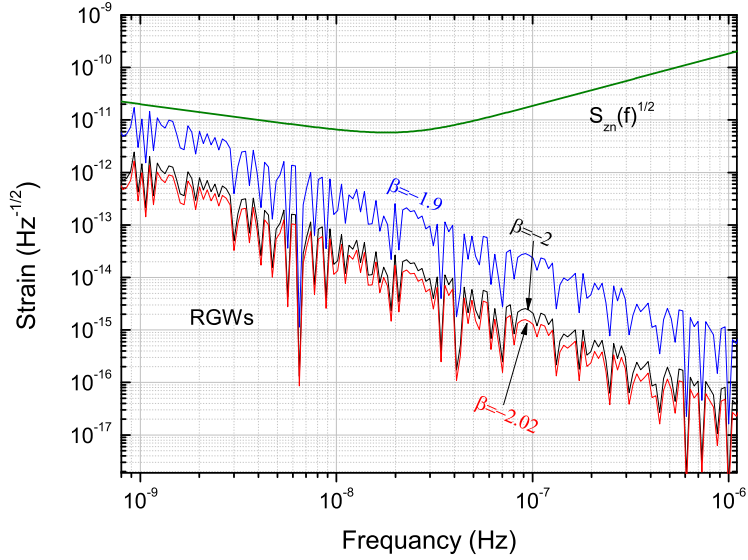}
   \caption{The strain sensitivity curve to be explored by the noise spectrum of the relative frequency fluctuations for a single pulsar timing. The quantities, $h_c(f)\sqrt{F/f}$,  of RGWs with  different parameters  $\beta=-1.9$, $\beta=-2.0$ and $\beta=-2.02$, respectively, are also demonstrated for comparison.}
   \label{Fig5}
   \end{figure}

The one-sided power spectrum of the induced timing residuals by RGWs, $P(f)$, is defined as (\citealt{Jenet+etal+2006})
\be\label{residual}
\int_0^\infty P(f)df=\sigma^2,
\ee
where $\sigma^2$ is the variance of the timing residuals generated by the stochastic background of RGWs.
The power spectrum $P(f)$ is related to the characteristic strain as
\be
P(f)=\frac{1}{12\pi}\frac{1}{f^3} h_c(f)^2.
\ee
$P(f)$ has the unit of $s^3$.  
 Fig. 6 shows the corresponding one-sided power spectrum of the induced timing residuals by RGWs with $\beta=-1.9$, $\beta=-2$ and $\beta=-2.02$, respectively.  It can be seen that RGWs with $\beta=-1.9$ leads to a higher $P(f)$ by about two orders of magnitude than that given by RGWs with $\beta=-2$. For $\beta=-1.9$, the power spectrum is as high as $10^{-5}$ $s^3$ around $10^{-9}$ Hz.
In fact, the integrating limits are determined by observational strategy. The lowest detectable frequency is given by $1/T$, where $T$ is the total time span of the data set. The highest one is determined by the Nyquist sampling rate. If one  observes pulsars  at   an interval of $\Delta t$, then the highest frequency is $2/\Delta t$.  For the observation of pulsar timing, $\Delta t$ is typically two weeks, and total data span are assumed to be around 10 years. 
Then, based on Eq.(\ref{residual}),  one has the standard deviation of the timing residuals $\sigma=3.2$ ns,  $0.4$ ns and $0.3$ ns for $\beta=-1.9$, $-2$ and $-2.02$, respectively.  For comparison, we also calculated the case of $\beta=-1.8$, $\sigma=26.2$ ns, even though $\beta$ is not possibly so large. 

\begin{figure}
   \centering
   \includegraphics[width=12.0cm, angle=0]{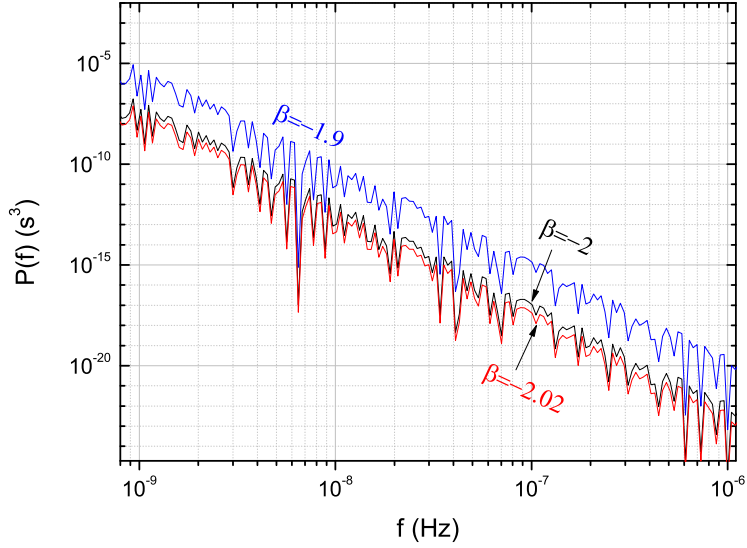}
   \caption{The one-sided power spectra of the  induced timing residuals by RGWs with different $\beta$.}
   \label{Fig6}
   \end{figure}

\section{Summary}

In summary, we analyzed the effects of RGWs on the pulsar timing residuals, based on some
reasonable parameters constrained by the recent observations of PPTA and the physical processes happened in the very early universe.
First of all, we corrected the normalization of RGWs by the CMB observations, and now the spectra are reduced  by two orders of magnitude compared to our previous results. Then, we compared the analytic spectrum and the approximate spectrum of RGWs, and we found they matched   each other very well. Therefore, one can  take advantage of  them alternatively. When we constrain the parameter  $\beta$ by the PPTA ,  the approximate spectrum is applied simply; while we calculated the 
total energy density contrast and the timing residuals induced by RGWs, the analytic spectrum is used. 
The current PPTA gives a constraint, $\beta<-1.70$, and the future PPTA would give a constraint as $\beta<-1.76$. 
On the other hand, the constraints of $\beta$ from the BBN/CMB    are dependent on some other cosmic parameters, such as the temperature of the reheating process $\trh$ and the expansion times of the reheating process $\zeta_1$.  The strongest constraint from BBN/CMB   is     $\beta<-1.83$. Note that, the constraints from BBN/CMB are a little bit overestimated because of the effects of the neutrino free-streaming (\citealt{Weinberg+2004}), the $e^{+}e^{-}$ annihilation and the  QCD transition (\citealt{Schwarz+1998}; \citealt{Wang+etal+2008}).
It is worth to point out that  the constraints of $\beta$ from the  PPTA are more  convincing since it is independent of other parameters. 
 Based on these constraints, we chose $\beta=-1.9$, $\beta=-2$ and $\beta=-2.02$, respectively, for demonstration. We compared the sensitivity curve  determined by the ground clock and a white noise of $100$ ns with the predicted RGWs. It was found that RGWs can not be detected by single pulsar timing at present, however, RGWs with frequencies as low as $10^{-9}$ Hz are very hopeful to be detected  if the intrinsic red noises were understood. Note that, the noise spectrum discussed in this paper is quite ideal, since many other noise components are not included. We quantitatively calculated the rms residuals induced by RGWs with different values of $\beta$, and found that the rms residuals are $\sigma=3.2$ ns,  $0.4$ ns and $0.3$ ns for $\beta=-1.9$, $-2$ and $-2.02$, respectively. Moreover, the rms residual  can be as much as $26.2$ ns for $\beta=-1.8$. Besides, we also showed the power spectra of the induced timing residuals by RGWs with different values of $\beta$.  
 
     RGWs are a very important and effective tool to exploit the knowledges of the very early universe. All the afore-mentioned constraints on RGWs  will help us to know the early universe more clearly. For example, the parameters of expansion times of the reheating process $\zeta_1$, the expansion times  $\zeta_s$ of the radiation-dominated stage and some other parameters  would be known better.   Even though, the quantum normalization  is not employed  here in order to give a  general result,  it should be considered elsewhere for a more complete discussion.
Moreover, RGWs act as a key role to connect the subject of  cosmology with pulsar timing observations.

\normalem
\begin{acknowledgements}This work was supported by the National Natural Science Foundation of
China (Grant Nos. 11103024,  11373028 and 11403030),  the Science and Technology Research Development Program of Shaanxi Province,  CAS "Light of West China" Program, and the Open Project of  Key Laboratory for Research in Galaxies and Cosmology, Chinese Academy of Sciences (Grant No. 14010205).   
\end{acknowledgements}

\bibliographystyle{raa}
\bibliography{bibtex}

\end{document}